\documentclass[twocolumn, superscriptaddress]{revtex4}
\pdfoutput=1
\usepackage{epsfig,graphicx}
\usepackage{amsmath,amssymb,amsfonts,amsthm}
\usepackage{bm} 
\usepackage{flafter}
\usepackage{color}
\usepackage{epstopdf}
\usepackage{latexsym}
\usepackage[colorlinks=true, citecolor=blue, urlcolor=blue ]{hyperref}
\usepackage{pxfonts,txfonts}
\usepackage{tgpagella}
\usepackage[T1]{fontenc}
\usepackage{ulem} 
\usepackage[utf8]{inputenc}
\newcommand{\bra}[1]{\langle #1 |}
\newcommand{\ket}[1]{| #1 \rangle}

\newcommand{\la}[1]{\lambda}
\graphicspath{{figures/}}

\usepackage{braket}

\begin{document}

\title{Entanglement scaling at first order phase transitions}
\author{A. Yuste}
\email{abel.yuste@uab.cat}
\affiliation{Departament de F\'{\i}sica, 
Universitat Aut\`{o}noma de Barcelona, 08193, Bellaterra, Spain.}
\author{C. Cartwright}
\affiliation{Centre for Theoretical Atomic, Molecular, and Optical Physics, School of Mathematics and Physics, Queen’s University Belfast, Belfast BT7 1NN, United Kingdom}
\author{G. De Chiara}
\affiliation{Centre for Theoretical Atomic, Molecular, and Optical Physics, School of Mathematics and Physics, Queen’s University Belfast, Belfast BT7 1NN, United Kingdom}
\author{A. Sanpera} 
\affiliation{Departament de F\'{\i}sica, 
Universitat Aut\`{o}noma de Barcelona, 08193, Bellaterra, Spain.}
\affiliation{ICREA, Pssg. Llu\'is Companys 23, 08010, Barcelona, Spain.}

\begin{abstract}
First order quantum phase transitions (1QPTs) are signaled, in the thermodynamic limit, by discontinuous changes in the ground state properties. These discontinuities affect expectation values of observables, including spatial correlations. When a 1QPT is crossed in the vicinity of a second order one (2QPT), due to the correlation length divergence of the latter, the corresponding ground state is modified and it becomes increasingly difficult to determine the order of the transition when the size of the system is finite. Here we show that, in such situations, it is possible to apply finite size scaling to entanglement measures, as it has recently been done for the order parameters and the energy gap, in order to recover the correct thermodynamic limit \cite{Campostrini14}. Such a finite size scaling can unambigously discriminate between first and second order phase transitions in the vicinity of multricritical points even when the singularities displayed by entanglement measures lead to controversial results.
\end{abstract}
\maketitle
\section{Introduction}
Understanding how many-body interacting systems order into different quantum phases as well as the transitions between them remains one of the most challenging open problems in modern physics. Quantum phase transitions (QPTs) are associated with the non-analytical behavior of some observable and/or correlator which can be either local or non local. With the discovery of topological and new exotic phases \cite{Kosterlitz73}, which fall outside Landau's symmetry breaking paradigm, the use of entanglement to describe quantum matter seems to be, by all means, necessary \cite{Amico08}. \\
Here, we restrict ourselves to a large class of QPTs which, in analogy to their classical counterparts, are signalled by singularities in the derivatives of the (free) energy \cite{Sachdev01}. In such cases, phase transitions are classified according to the minimum order of the derivative of the ground state energy which is not continuous. Correspondingly, 1QPTs (2QPTs) show singular behavior on the first (second) derivative of the ground state energy. For 1QPTs, the singular behavior translates into an abrupt discontinuity of some local observables while for 2QPTs the order parameters change continuously with a power law scaling. This scaling is characterized by the critical exponents allowing the classification of apparently different 2QPTs into the same universality class. 
With such definitions at hand it looks straightforward to distinguish if a given QPT is of first or second order. But this is not always the case when finite size effects are present. This question becomes especially relevant in the vicinity of multicritical points where several QPTs of different order coexist in a narrow range of Hamiltonian parameters.

First studies of the entanglement behavior close to a QPT were performed by analyzing bipartite entanglement in simple spin models like the quantum Ising spin-1/2 chain which exhibits a 2QPT. In \cite{Osterloh02, Osborne02}, the authors show that, similarly to the ground state energy behavior, the bipartite entanglement between two adjecent spins, as measured by the concurrence ($\mathcal{C}$), is a continuous function of the parameter driving the transition but its first derivative is singular at the critical point in the thermodynamical limit. Moreover, they showed that, for finite size systems, a finite size scaling (FSS) can be performed on the concurrence allowing the correct extraction of the critical exponents corresponding to the Ising transition. These seminal contributions opened a new path in the analysis of QPTs using entanglement. Since then, a large amount of work has been devoted to deepen the connections between quantum information and QPTs, see for instance \cite{Bose02,GVidal03,Latorre04,JVidal04,Alcaraz04,JVidal04b,Barnum04,Orus10,Gu13,Stasinska14,Guhne14}.
All these previous results were brought to a more general form in~\cite{Lidar04} using the Kohn theorem  which links the ground state energy of a local Hamiltonian, $\hat{H}(\lambda)=\sum_k \hat{H}_k(\lambda)$, where each $\hat{H}_k$ involves as a maximum $k$-body interactions, with the corresponding $k$-reduced density matrices. In the usual case of local Hamiltonians involving just two-body interactions of two particles $i$ and $j$, $\hat{H}_{ij}$,  the energy of the ground state $\ket{\Psi_0}$ can be written as
\begin{equation}
E_0(\lambda)=\sum_{i,j} {\rm Tr}\;\left(\hat{H}_{ij}\;\rho^{ij}\right),
\end{equation}
where 
$\rho^{ij}$ is the corresponding reduced density matrix.
If the local Hamiltonians, $\hat{H}_{ij}$, depend smoothly on the parameter $\lambda$, then a one-to-one correspondence can be made between the singularities of $E_0(\lambda)$ arising in a QPT and the singularities  of (the matrix elements) of $\rho^{ij}$. In turn, this fact translates into singularities of entanglement measures, like the concurrence, which depends exclusively on $\rho^{ij}$. Therefore, if $\hat{H}_{ij}$ are smooth functions of the $\lambda$ parameter that drives the phase transition, it generally follows that $\partial_{\lambda} E_0\sim \rho^{ij}$. In other words, for a first order quantum phase transition, the singularity on the ground state energy should arise from a singularity in $\rho^{ij}$. As a consequence, the singularity should be evident in any bipartite entanglement measures depending on the reduced density matrix, $\rho^{ij}$, for example in the concurrence. By the same reasoning, since a 2QPT has a singularity in the second derivative of the ground state energy and $\partial^{2}_{\lambda}E_0 \sim\partial_{\lambda}\;\rho^{ij}$,  it follows that a 2QPT should display a singularity in $\partial_{\lambda}\mathcal{C}$. This was nicely illustrated in~\cite{Osterloh02,Osborne02},  demonstrating the fact that quantum phase transitions are related to entanglement.  
The above results were cast into a theorem in~\cite{Lidar04} stating that, unless there exist accidental divergences, the order and properties (e.g. critical exponents, critical point) of first and second order quantum phase transitions are signaled by singularities on entanglement measures depending on the appropriate reduced density matrix of the corresponding ground state. The theorem works in both directions, i.e., a discontinuity in e.g. the concurrence signals a 1QPT while a discontinuity/divergence in its derivative signals a 2QPT.\\
The above theorem has some known caveats. For instance, entanglement measures often involve optimization procedures which may lead to accidental non-analytic behavior. In the spin-1/2 XXZ model, for example, there is a continuous $\mathcal{C}$ and a discontinuous derivative along the 1QPT between the ferromagnetic and critical phase~\cite{Syljuasen03}. 
For the same transition, it has been shown that a symmetry breaking in the ferromagnetic phase also modifies the origin of the non-analytic behavior of the concurrence ~\cite{Oliveira14}. Also, a non-analytic $\mathcal{C}$  in the absence of a QPT for a model with three body interactions has been reported in \cite{Yang05}.\\
Here, we analyze scaling properties of bipartite entanglement measures near multicritical points. Although finite size scaling is a tool to obtain the thermodynamical properties of the system for continuous phase transitions, here we show that such a tool can be employed also in entanglement measures for 1QPTs. Further we demonstrate that when finite size effects are important,  it is precisely the scaling of the entanglement measure and not the measure itself which determines the correct order of the transition. This fact is especially relevant for 1QPT crossed in the vicinity of 2QPT and it is in accordance with the recent results reported by Campostrini and coauthors \cite{Campostrini14} showing that the order parameter of a 1QPT can be continuous for finite systems and admits an appropriate finite size scaling. \\
The paper is organized as follows. In Sec. II we report how the concurrence scales in the spin-1/2 Ising chain in transverse and longitudinal field. In Sec. III we analyze the negativity in a spin-1 model with several 1QPTs. Finally, in Sec. IV we discuss the results and conclude. 

 
\section{Spin-1/2 Ising model with longitudinal field}
\label{Section:Spin1/2}
\begin{figure}
\includegraphics[scale=0.4]{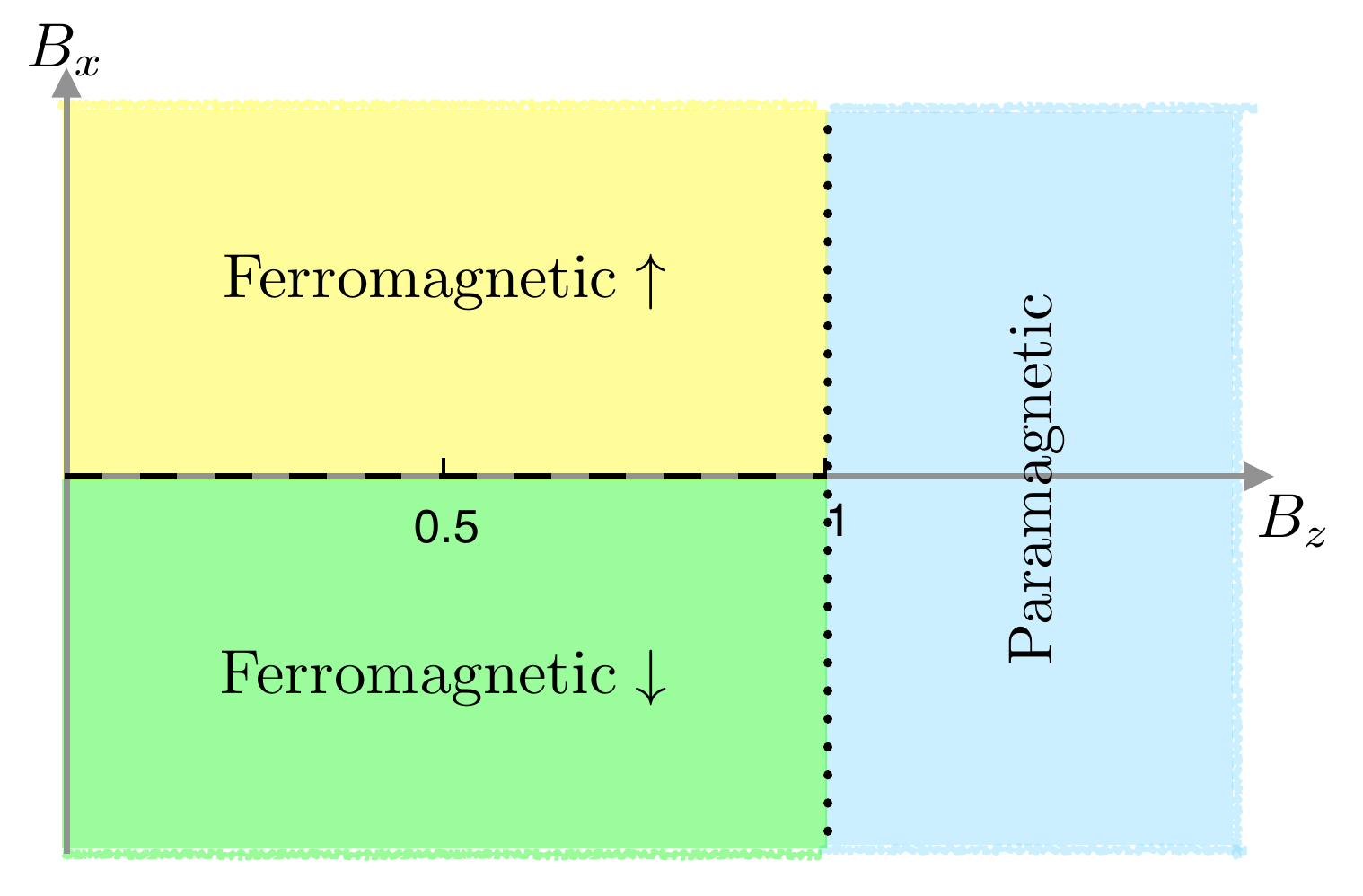}
\caption{Phase diagram for the spin-1/2 Ising model with a longitudinal field. The dashed line depicts the 1QPT while the dotted line the 2QPT. }
\label{fig:phase_spin1/2}
\end{figure}

\begin{figure*}
\includegraphics[width=1.9\columnwidth]{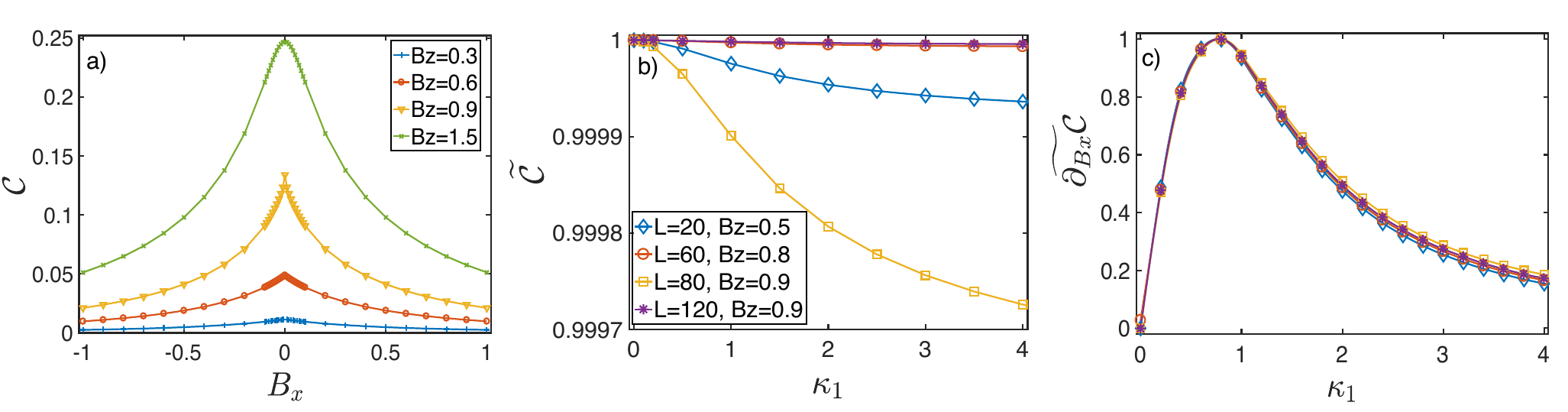}
\caption{Concurrence near the 1OQPT of the Ising model with longitudinal field, Eq. \ref{eq:Ham1/2}. Panel a), concurrence for L=40. We observe a spike at the 1QPT critical point $Bx=0$ for $B_z<0$. Panel b), concurrence normalized by its maximum, $ \widetilde{\mathcal{C}}$, as a function of the scaling variable $\kappa_1$ defined in \eqref{Eq:kappa_ising}, there is no data collapse. Panel c), derivative of the concurrence normalized by its minimum,  $\widetilde{\partial_{Bx}\mathcal{C}}$ , plotted as a function of $\kappa_1$ showing a universal behavior for the same set of values than in panel b).}
\label{fig:concu}
\end{figure*}

The first spin model we explore is the spin-1/2 Ising model with a longitudinal field, 
\begin{align}
	\label{eq:Ham1/2}
\hat{H}=-J\sum_{i=1}^{L-1}\hat{\sigma}_{i}^{x}\hat{\sigma}_{i+1}^{x}-B_z\sum_{i=1}^{L}\hat{\sigma}_i^z-B_x\sum_{i=1}^L\hat{\sigma}_i^x
\end{align}
where  $\hat{\sigma}_{i}^\alpha$ are the Pauli matrices for spin $i$ and we set $J=1$ and $B_z\geq0$. In Fig.~\ref{fig:phase_spin1/2}, we provide the phase diagram of the model. For $B_x=0$, where the system reduces to the integrable Ising model, there is a 2QPT at $B_z=1$ between the ferromagnetic ($B_z<1$) and paramagnetic phases ($B_z>1$). This 2QPT was the first one studied by means of bipartite entanglement, \cite{Osterloh02}. When $B_x\neq0$, the system is no longer integrable and we use the Density Matrix Renormalization Group (DMRG) with open boundary conditions (OBC) \cite{White92, White93, DeChiara06} and exact diagonalization (ED) calculations. When the system is in the ferromagnetic phase, a 1QPT takes place at $B_x=0$ between the two ferromagnetic ground states, ferromagnetic $\uparrow$ and ferromagnetic $\downarrow$. This transition can be detected with a discontinuity in the magnetization, $M_x=\sum_i\langle\hat{\sigma}^x_i\rangle$, which passes from positive to negative values.
For finite systems, numerical calculations in a region sufficiently close to $B_x=0$ lead to a smooth slope in $M_x$ instead of a discontinuity. To deal with this effect, in Ref.\cite{Campostrini14} a FSS is proposed for transitions driven by a magnetic field, $h$,  with the argument that the relevant  scaling variable, $\kappa$, should depend on the ratio between the energy contribution of $h$, and the gap at the critical point, $\Delta_L$,
\begin{align}
\label{Eq:kappa_general}
\kappa\sim\dfrac{hL}{\Delta_L}.
\end{align} 
This quantity can be interpreted as a zoom in the $h$ range where the change in the system energy is comparable to $\Delta_L$. When $hL\sim\Delta_{L}$, quantum fluctuations mix both ground states and avoid the energy crossing. As a result, the ground state is continuous along the transition and so is the order parameter associated to the transition. This feature is obviously stronger as we get closer to a 2QPT where the correlation length of the system ($\xi$) diverges, enhancing the nearby finite size effects and thus increasing $\Delta_{L}$. As we get away from the 2QPT, $\xi$ decreases, $\Delta_{L}\rightarrow 0$, and the relevant range of $h$ shrinks, effectively inducing a discontinuity even for small systems. In \cite{Campostrini14,Campostrini15}, it is shown that both, the first energy gap and the magnetization obey the scaling ansatz,
\begin{align}
\label{Eq:scaling_gap}
\Delta(L,B_x)\approx\Delta_Lf_{\Delta}(\kappa)
\end{align}
\begin{align}
\label{Eq:scaling_magne}
M_x(L,B_x)\approx m_0f_M(\kappa)
\end{align}
where $f_{\Delta}(\kappa)$ and $f_M(\kappa)$ are continuous universal functions for all $L$ and $B_z$. For the  model in \eqref{eq:Ham1/2}, which is integrable for $B_x=0$, the scaling variable can be defined as \cite{Campostrini14},
 \begin{align}
\kappa_1=\dfrac{2m_0B_xL}{\Delta_{L}},
\label{Eq:kappa_ising}
\end{align}
where $\Delta_{L}\approx 2(1-B_z^2)B_z^L$ is the first gap for OBC at the critical point, $B_x=0$, and 
\begin{align}
\label{Eq:m0}
m_0=\lim_{h\to 0^+}\lim_{L\to \infty}\langle \hat{\sigma}_x\rangle=(1-B_z^2)^{1/8}.
\end{align}

Here we apply the concepts developed in Ref.~\cite{Campostrini14} to study how entanglement behaves across the same transition. We use, as a measure of entanglement, the concurrence \cite{Wooters97},
\begin{align}
\mathcal{C}={\rm max}(0,\lambda_1-\lambda_2-\lambda_3-\lambda_4),
\end{align}
where $\lambda_i$ are the eigenvalues, in decreasing order, of $R=\sqrt{\sqrt{\rho}\tilde{\rho}\sqrt{\rho}}$ with $\tilde{\rho}=(\hat{\sigma}_y\otimes\hat{\sigma}_y)\rho^*(\hat{\sigma}_y\otimes\hat{\sigma}_y)$.
Since $\mathcal{C}$ is supposed to be discontinuous at the critical point for 1QPTs, naively we might expect that it will follow a similar scaling behavior as $M_x$, Eq.~\eqref{Eq:scaling_magne}. In Fig.~\ref{fig:concu}, we show  $\mathcal{C}$ for the two central spins which, if border effects are neglected,  will also hold for the rest of neighboring spin pairs. 
In the left panel, we observe that $\mathcal{C}$ is a continuous function with a spike at the critical point which signals a singularity in the first derivative, $\partial_\lambda \mathcal{C}$. The spike is more pronounced as the transition is closer to the 2QPT at $B_z=1$. It is worth pointing out that such behavior bears strong similarities with the  geometric entanglement, a collective measure of entanglement indicating how much the ground state differs from a separable state~\cite{Orus10}. Notice that for $B_{z} > 1$, when the system is in the paramagnetic phase independently of the sign of the longitudinal field $B_{x}$, (as indicated in Fig.~\ref{fig:concu} for $B_{z}=1.5$) , the concurrence becomes an analytical function of $B_{x}$. In the central panel of Fig.~\ref{fig:concu}, we plot the scaling of the concurrence normalized by its maximum value,  $\widetilde{\mathcal{C}}=\mathcal{C}/\mathcal{C}_{max}$, as a function of the scaling variable $\kappa_1$ to determine whether $\mathcal{C}$ fulfills a similar scaling relation as in  Eq.~\eqref{Eq:scaling_magne}. It is enough to investigate what happens for $\kappa_1>0$ because $\mathcal{C}(\kappa_1)$ has even parity, i.e., $\mathcal{C}(\kappa_1)=\mathcal{C}(-\kappa_1)$.  Finally, in panel c) we display the scaling of the derivative of the concurrence normalized by its minimum value $\widetilde{\partial_{Bx}\mathcal{C}}=\partial_{Bx}\mathcal{C}/[\partial_{Bx}\mathcal{C}]_{min}$. Interestingly enough, the concurrence does not scale with the fitting parameter $\kappa_1$, but its derivative does.
In the central panel, we can see that the data for different $B_z$ and $L$ does not collapse in a universal function, while, in the right panel, there is a good data collapse for different values of $B_z$ and $L$. Thus, $\partial_{Bx}\mathcal{C}$ fulfills the scaling anstatz,  
\begin{align}
\partial_{Bx}\mathcal{C}(L,B_z)=[\partial_{Bx}\mathcal{C}]_{min}g(\kappa_1)
\label{Eq:scaling_C}
\end{align}
where $g(\kappa_1)$ is a universal function for any $L$ and $B_z$. We further discuss all these results in Sec.~\ref{Section:Conclusions}.
\section{spin-1 XXZ model with on-site anisotropy}
\label{Section:Spin1}
\begin{figure}
\includegraphics[scale=0.5]{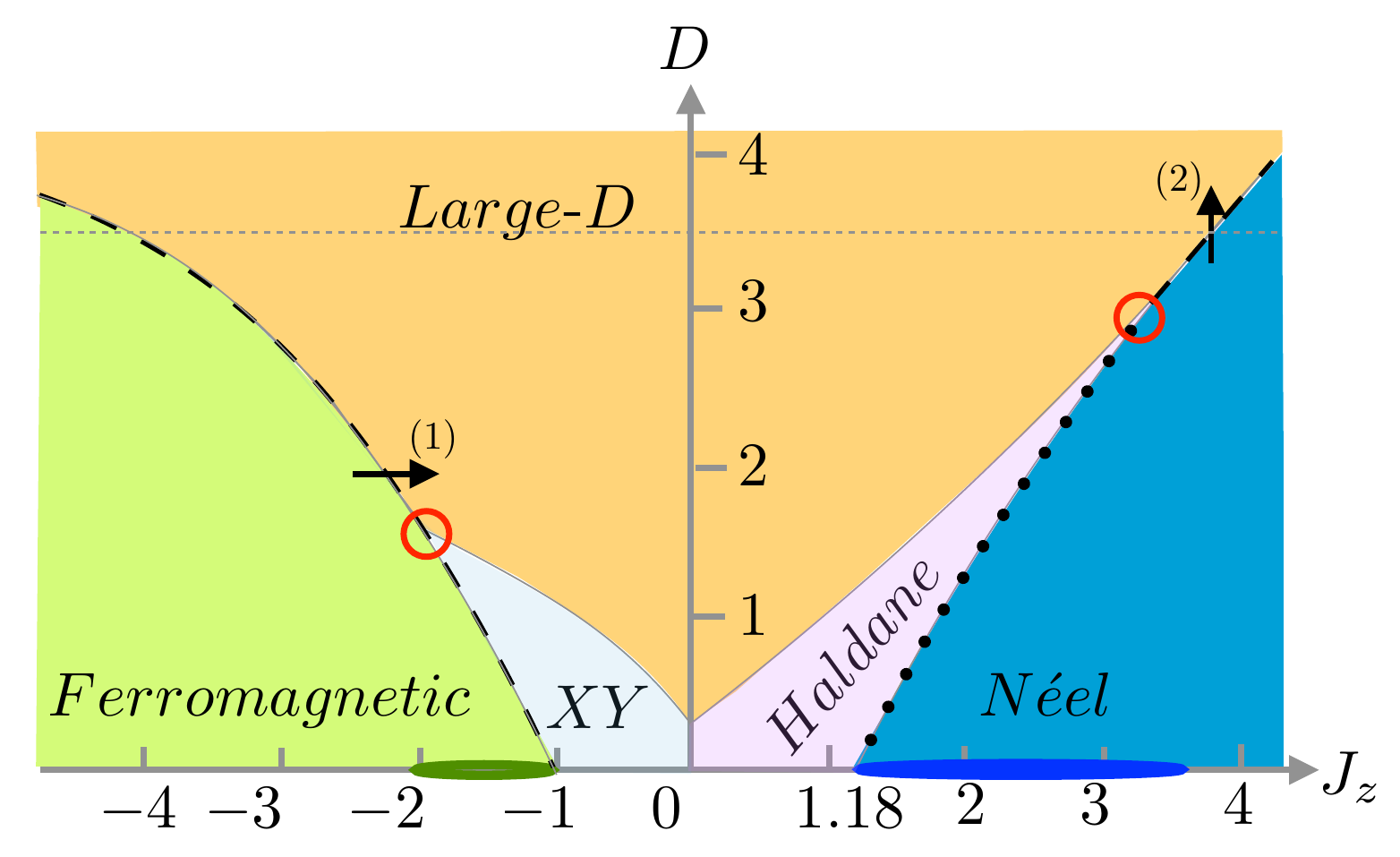}
\caption{Phase diagram for the spin-1 model in \eqref{eq:Ham1} with $D>0$. The dashed lines depict 1QPTs and the arrows the points where we cross them. The black dotted line depicts the 2QPT between Haldane and N\'eel phases. The ovals show the areas where we add an external field to look into the 1QPT between the two-fold degenerated ground states in the ferromagnetic and N\'eel phases. The red circles signal the tricritical points present in the phase diagram.}
\label{fig:phase_spin1}
\end{figure}
In this section, we extend the study of entanglement along 1QPTs to a spin-1 system. Since $\mathcal{C}$ cannot be used to measure the entanglement of two spin-1 particles, we use the negativity, a lower bound of the actual entanglement \cite{Zyczkowski98, Vidal02}:
\begin{align}
\mathcal{N}=\dfrac{||\rho^{T_B}||_1-1}{2},
\end{align}
where $||\rho^{T_B}||_1$ is the sum of the absolute value of all singular values of the partially transposed density matrix.
The model under scrutinity is the spin-1 XXZ with on-site anisotropy,
\begin{align}
	\label{eq:Ham1}
\hat{H}=\sum_{l=1}^{L-1}\left[\hat{S}^x_l\hat{S}^x_{l+1}+\hat{S}^y_l\hat{S}^y_{l+1}+J_z\hat{S}^z_l\hat{S}^z_{l+1}\right]+D\sum_{l=1}^{L}\left(\hat{S}^z_l\right)^2,
\end{align}
where $\hat{S}_l^\alpha$ are the spin-1  matrices for spin $l$ and $D$ is the uniaxial single-ion anisotropy which we take as positive. We choose this model because of the richness of its phase diagram \cite{Sanctuary}, schematically shown in Fig.~\ref{fig:phase_spin1}, with several 1QPTs which we depict with dashed lines. 
\begin{figure}
\includegraphics[width=0.7\columnwidth]{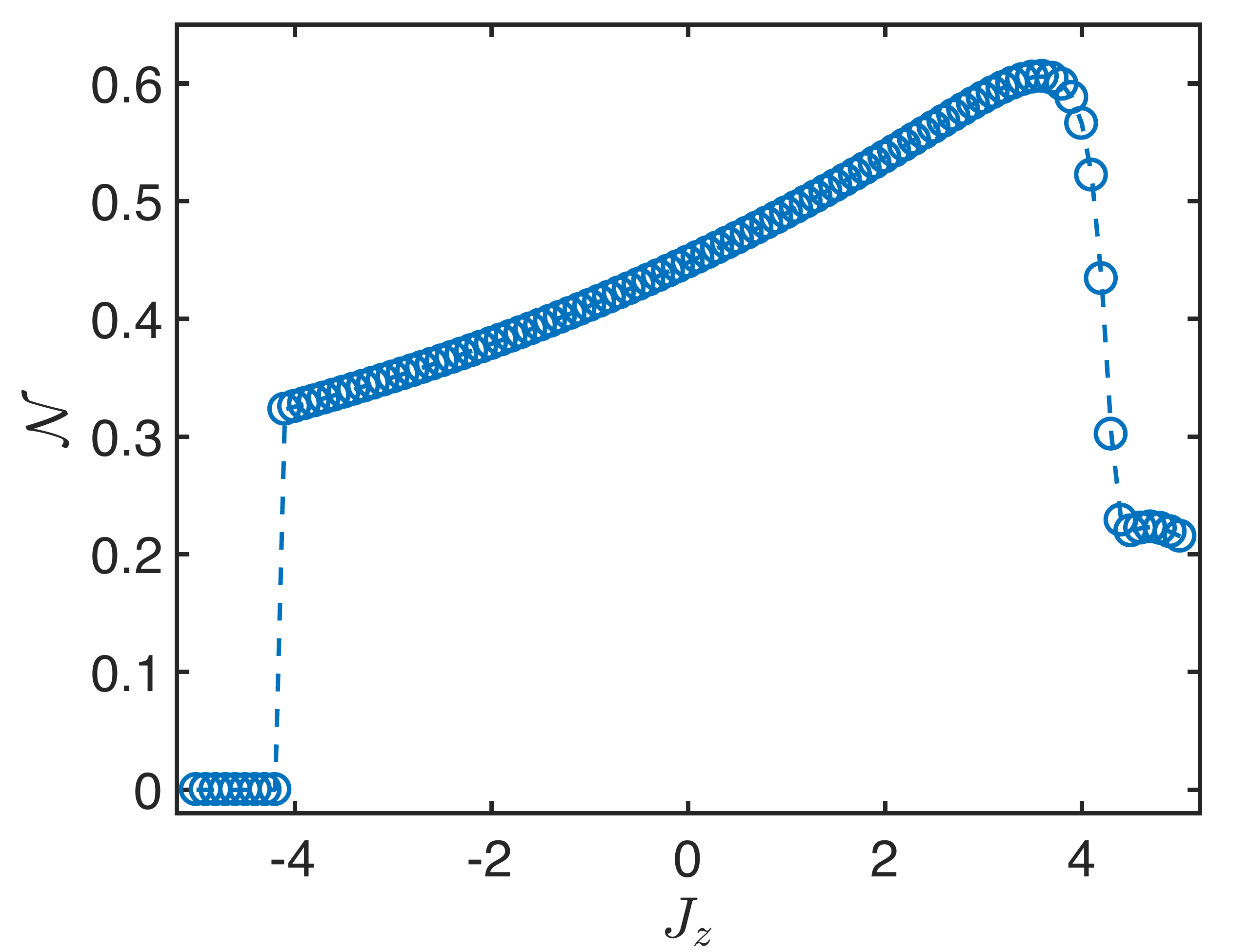}
\caption{Negativity as a function of $J_z$ for $D=3.5$ and $L=8$. We observe a discontinuity in the 1QPT from ferromagnetic to large $D$ phases and a smooth slope for the 1QPT between the large $D$ and N\'eel phases. }
\label{fig:D35}
\end{figure}
The behaviour of $\mathcal{N}$  along the different 1QPTs present in the model can be very different depending on their closeness to a multicritical  point which also involves 2QPTs. We start by examining the negativity as a function of the $J_z$ for a constant uniaxial field $D=3.5$ and $L=8$. Two 1QPTs are crossed at such value of $D$ as indicated in the phase diagram by a grey horizontal line (see Fig.~\ref{fig:phase_spin1}). The first one corresponds to the transition from ferromagnetic order to the large $D$  phase, which is crossed approximately at $J_z=-4.2$. Another 1QPT appears between large $D$/N\'eel at approx $J_z=3.8$. As clearly shown in Fig.~\ref{fig:D35}, the former phase transition is clearly signalled by a discontinuity in $\mathcal{N}$, whereas the latter shows a smooth slope along the transition.

\begin{figure}
\includegraphics[width=\columnwidth]{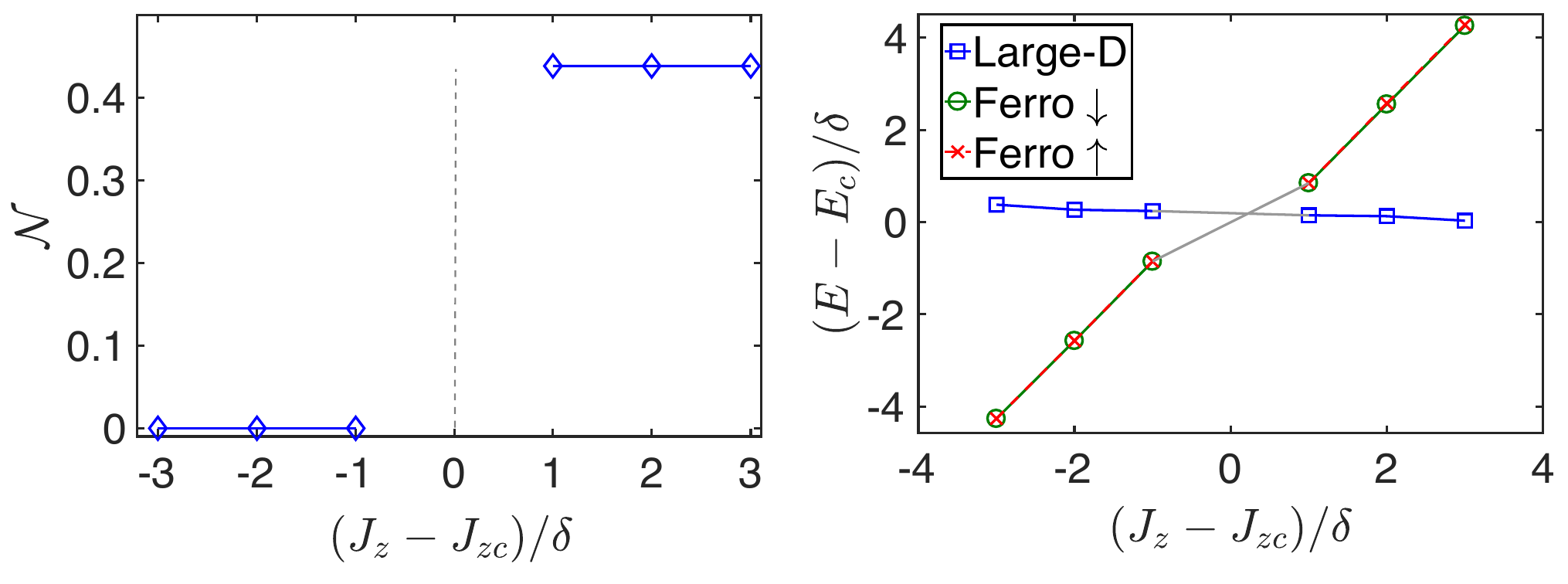}
\caption{Phase transition between ferromagnetic and large $D$ for $D=2$ and $L=8$. Left panel, we observe a jump in the negativity (dotted line) for $J_{zc}$ even when using a step $\delta=10^{-13}$. Right panel, the corresponding energy crossing between the two-fold degenerated ferromagnetic ground state at $J_z<J_{zc}$ (circles and crosses) and the large $D$ ground state at $J_z>J_{zc}$ (squares).}
\label{fig:Ferro_largeD}
\end{figure}

In Fig.~\ref{fig:Ferro_largeD}, we show in more detail the behavior of $\mathcal{N}$ along the ferromagnetic/large $D$ phase transition.  This transition is depicted in Fig.~\ref{fig:phase_spin1} by number (1). The results, in this case, are for $D=2$ and show that the behavior is the same along all the transition line, even if we are close to the XY phase. In the left panel, we show $\mathcal{N}$ very close to the phase transition critical point, $J_{zc}$, with a very small step in $J_z$ of $\delta=10^{-13}$. We can observe that $\mathcal{N}$ is still discontinuous even with such a high precision  around $J_{zc}$. In the right panel, we show that there is a neat level crossing between the two-fold degenerated ferromagnetic ground state, for $J_z<J_{zc}$, and the large $D$ ground state, for $J_z>J_{zc}$. This means that, at $J_{zc}$, there is a sudden change in the ground state which we can detect with a discontinuity in $\mathcal{N}$. Therefore, in Fig.~\ref{fig:Ferro_largeD}, we observe the expected discontinuous behavior for $\mathcal{N}$ along a 1QPT even for a system of just 8 spins without any finite size effects. \\
\begin{figure}
\includegraphics[width=0.9\columnwidth]{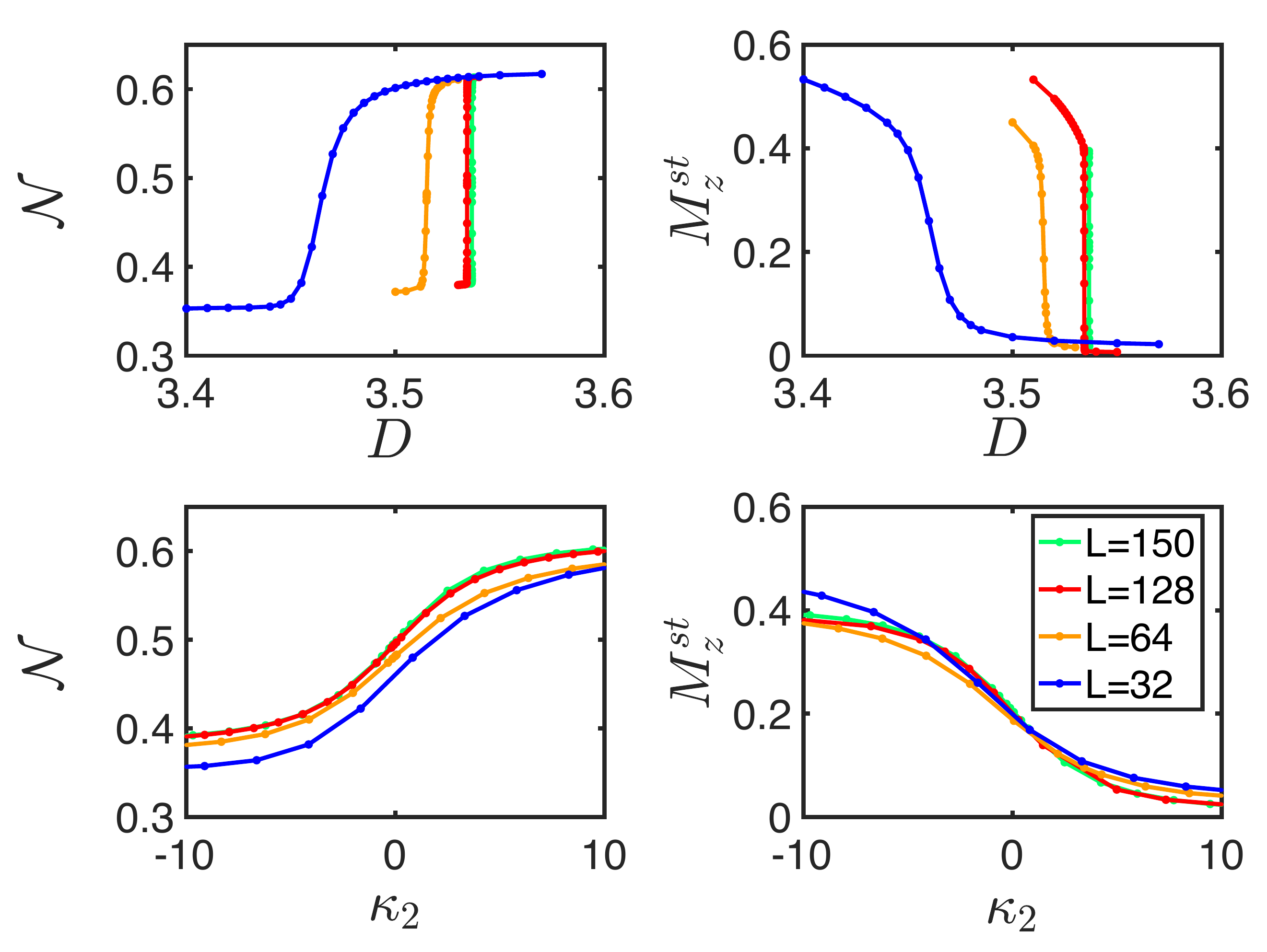}
\caption{Left column: negativity. Right column: staggered magnetization. Results are for $J_z=3.8$ and different $L$.  In the first row, we plot the quantities as a function of $D$. We observe a smooth slope of both the negativity and the staggered magnetization where the 1QPT is expected. Bottom row, quantites plotted as a function of $\kappa_2$ showing the tendency to converge towards a universal function}
\label{fig:LargeD_N\'eel}
\end{figure}
In Fig.~\ref{fig:D35}, though, we show that a smooth slope in $\mathcal{N}$ rather than a discontinuity is observed for the the large $D$/N\'eel transition. In Fig.~\ref{fig:LargeD_N\'eel}, we study with more detail this transition fixing $J_z=3.8$ and varying $D$ around the critical point $D_{c,L}$ depicted by (2) in Fig.~\ref{fig:phase_spin1}. We add the subindex $L$ to indicate that this quantity now depends on the system size. In order to show the behavior for larger systems, we have used DMRG calculations for $L=32, 64, 128$ and $150$. In the two top panels, we observe that both, $\mathcal{N}$ and the staggered magnetization, $M_z^{st}=\sum_{i=1}^{L}(-1)^{i}\langle \hat{S}_i^{z}\rangle$, change smoothly around the transition point. As $L$ is increased, the slope is more pronounced getting closer to a discontinuity and we need values of $D$ closer to the critical point to observe the continuous slope. For instance, the necessary step in $D$ to observe a continuous behavior is $\delta \sim10^{-4}$ and $\delta \sim10^{-6}$ for $L=64$ and $L=150$, respectively. Note that in this case, as in Sec.~\ref{Section:Spin1/2}, we are very close to a 2QPT. As we get further from it, for $J_z\gg 4$, this effect becomes less important and both $\mathcal{N}$ and $M_z^{st}$ are effectively discontinuous. 
Since the transition is known to be of first order, we propose a similar FSS as in the previous section, defining a relevant scaling variable, $\kappa_2$, as the ratio between the energy contribution of $D$ along the transition and the gap at the critical point,
\begin{align}
\label{Eq:kappa_2}
\kappa_2\sim\dfrac{(D-D_{c,L})L}{\Delta_L},
\end{align}
where now, $\Delta_L$ is obtained numerically and $(D-D_{c,L})L$ is a bare estimation for the energy contribution of the parameter $D$.
In the bottom panels, we plot $\mathcal{N}$ and $M_z^{st}$ as a function of this scaling variable $\kappa_2$. As we can observe, both quantities seem to converge, though not perfectly, towards a universal scaling, as described by \eqref{Eq:scaling_magne}. It is worth mentioning that \eqref{Eq:kappa_2} is an approximation whereas, in the previous section we had analytic expressions for $\Delta_L$ and $m_0$ in \eqref{Eq:kappa_ising}. Actually, in Ref.~\cite{Campostrini15}, a similar FSS, with non-analytic expressions, is proposed for the Potts chain with a similar convergence. It seems reasonable, thus, to state that for this 1QPT, when we are close to the 2QPT, $\mathcal{N}$ is continuous due to finite size effects and that it obeys the scaling ansatz for 1QPT.

\begin{figure*}
\includegraphics[width=1.9\columnwidth]{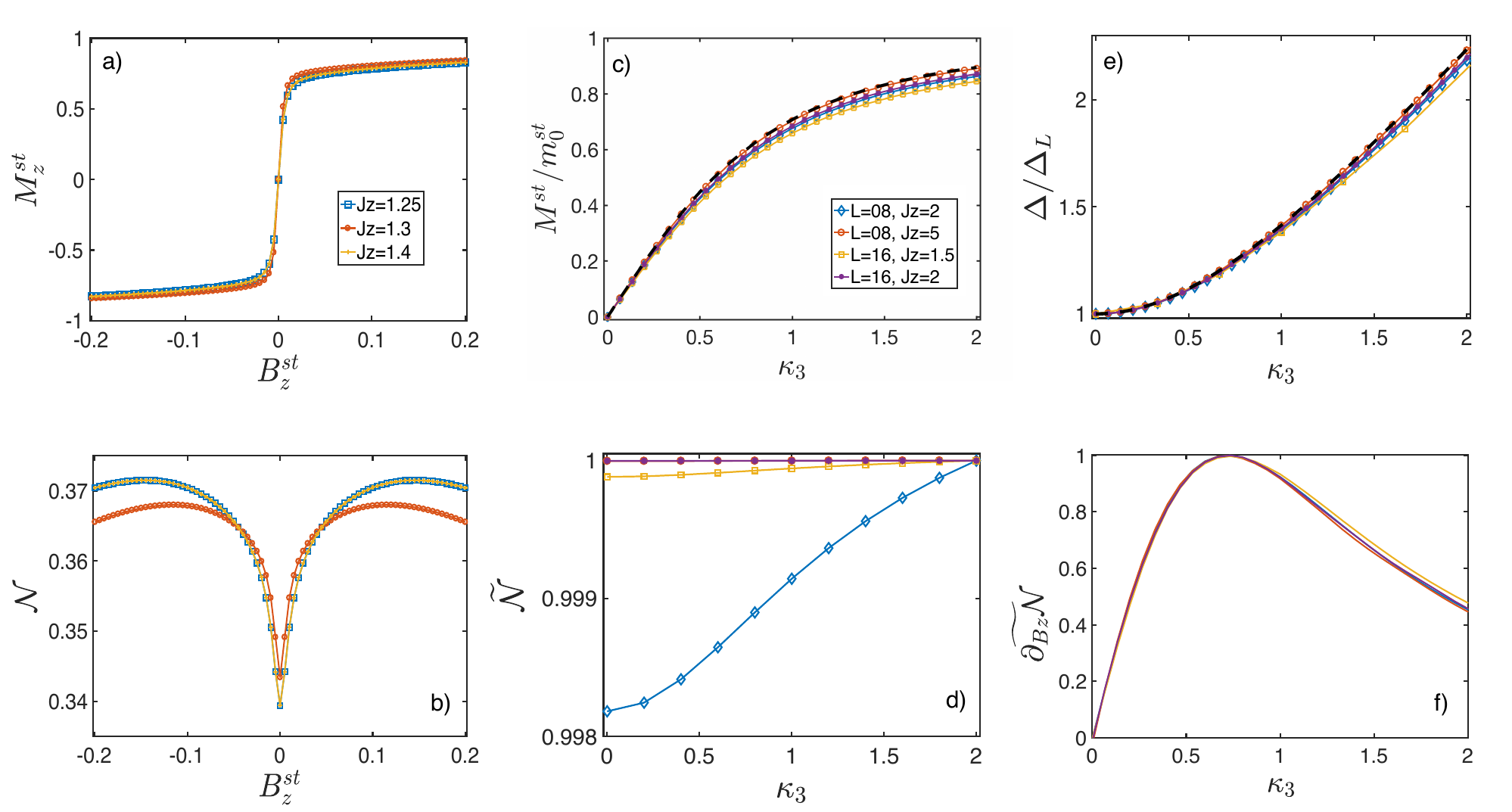}
\caption{Panels a) and b), $M_z^{st}$ and $\mathcal{N}$ as a function of $B_z^{st}$ for different $J_z$ and $L=32$ ($D=0$ in all the panels). Panels c) and d), $M_z^{st}/m_0^{st}$ and $\widetilde{\mathcal{N}}=\mathcal{N}/\mathcal{N}(B_z^{st}=0)$ plotted against the scaling variable $\kappa_3$, \eqref{Eq:kappa3}, (same legend holds for panels c), d), e) and f)). $M_z^{st}$ fulfils the universal scaling,  \eqref{Eq:scaling_magne} but  $\mathcal{N}$ does not. Panel e) and f), both $\Delta(\kappa_3)/\Delta_L$ and $\widetilde{\partial_{B_z^{st}}\mathcal{N}}(\kappa_3)$ (derivative normalized by its maximum) show, as well, a good data collapse. Dashed lines in panels c) and e) are analyticaly obtained in Ref.~\cite{Campostrini14}.}
\label{fig:N\'eel_AntiN\'eel}
\end{figure*}

Finally, we want to study $\mathcal{N}$ in a similar case as in the previous section for the spin-1/2 Ising model with transverse field, i.e., a 1QPT which is a consequence of a discrete $\mathbb{Z}_2$ symmetry breaking of a two fold-degenerated ground state. In the spin-1 model in \eqref{eq:Ham1}, this can be done by adding a magnetic field, $B_z\sum_{i=1}^L\hat{S}_i^z$,  in the ferromagnetic phase leading to a ferromagnetic $\uparrow$/ferromagnetic $\downarrow$ transition, or a staggered field, $B_z^{st}\sum_{i=1}^{L}(-1)^{i}\hat{S}_i^{z}$, in the N\'eel phase leading to a N\'eel/Anti-N\'eel transition. These regions are highlighted by the ovals along $D=0$ in Fig.~\ref{fig:phase_spin1}. The order parameter of both transitions, $M_z$ and $M_z^{st}$, respectively, are discontinuous in the thermodynamic limit. For the ferromagnetic phase the entanglement remains always constant and zero.
More interesting features appear in the N\'eel/Anti-N\'eel transition whose results are summarized in Fig.~\ref{fig:N\'eel_AntiN\'eel}. In panel a) and b) we show, respectively, $M_z^{st}$ and $\mathcal{N}$ as a function of $B_z^{st}$. While $M_z^{st}$ has the expected discontinuous behavior, $\mathcal{N}$ has a dip at the critical point, signalling a continuous $\mathcal{N}$ but a discontinuous derivative.
In order to apply a similar FSS ansatz that was used in Sec.~\ref{Section:Spin1/2}, we define as the relevant scaling variable,
\begin{align}
\label{Eq:kappa3}
\kappa_3=\dfrac{2m^{st}_0B_z^{st}L}{\Delta_L}.
\end{align} 
Since the 2QPT between Haldane/N\'eel, which is depicted by dotted lines in Fig.~\ref{fig:phase_spin1}, is known to be of the same universality class as the spin-1/2 Ising, we can expect that the analytic function used in \eqref{Eq:m0} is valid now for $m_0^{st}$ with the mapping, $B_z\rightarrow \left(\dfrac{J_{zc}}{J_z}\right)$, where $J_{zc}\approx 1.186$ is the 2QPT critical point for $D=0$, \cite{Ejima15}.
In panel c) and d) of Fig.~\ref{fig:N\'eel_AntiN\'eel}, we show, respectively, $M_z^{st}$ and $\mathcal{N}$ as a function of $\kappa_3$.  The scaling ansatz, \eqref{Eq:scaling_magne}, works fine for $M_z^{st}$, but, as it happened for $\mathcal{C}$ in the spin-1/2 Ising model, it fails for $\mathcal{N}$. In panels  e) and f) we show how, $\Delta_L$ fulfills the scaling ansatz, \eqref{Eq:scaling_gap},  and that the derivative of the negativity, $\partial_{B_z^{st}}N$, also verifies the scaling ansatz, \eqref{Eq:scaling_C}, which was fulfilled by $\partial_{B_x}\mathcal{C}$ in the previous section. To check the correctness of our results, in Fig.~\ref{fig:N\'eel_AntiN\'eel} we compare the numerical data for $M^{st}(\kappa_3)$ and $\Delta_L(\kappa_3)$ the analytic functions derived in Ref.~\cite{Campostrini14} as a result of a two level theory.
\section{Discussion and conclusions}
\label{Section:Conclusions}
\begin{figure}
\includegraphics[width=\columnwidth]{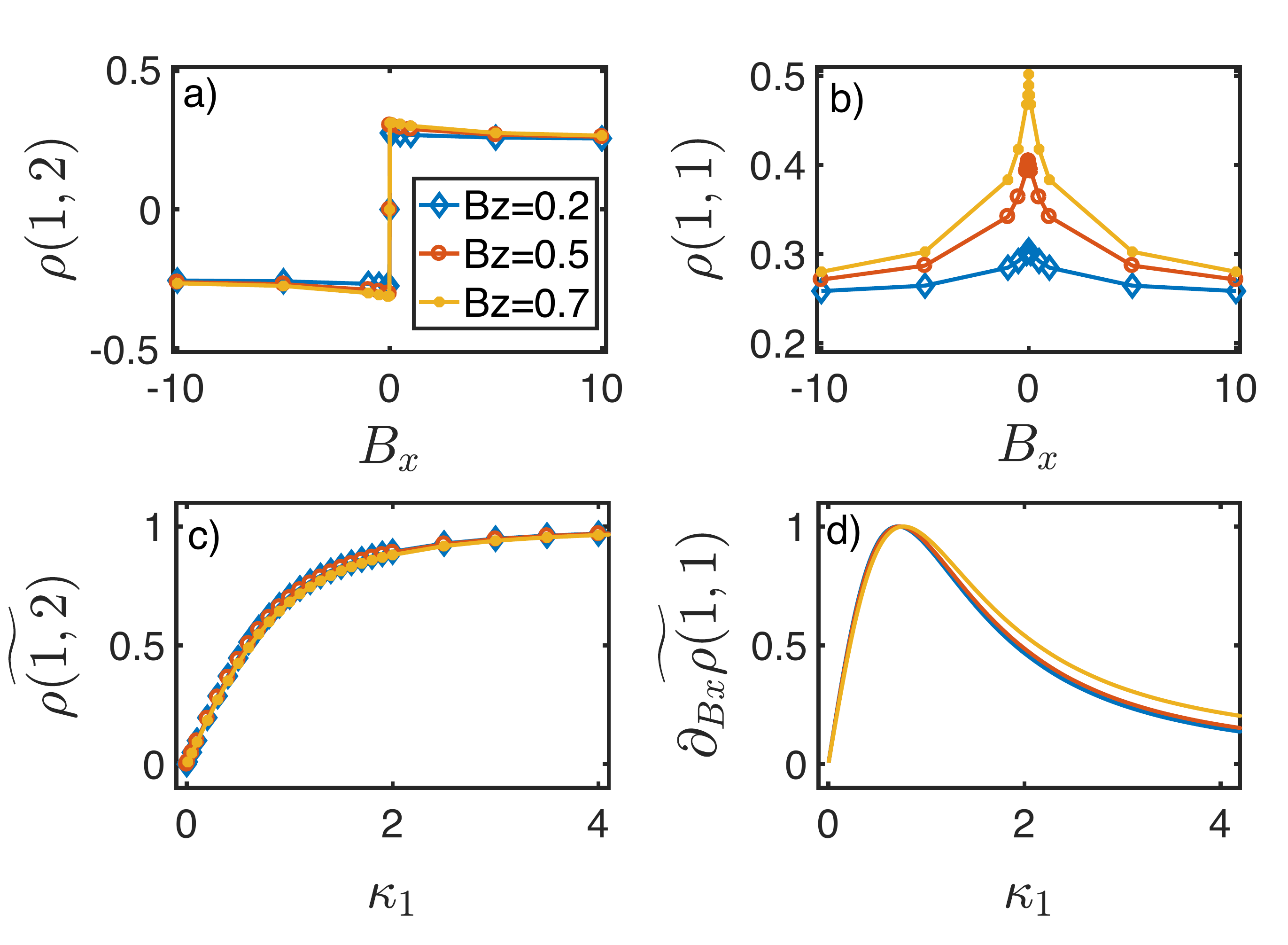}
\caption{Two of the elements of $\rho^{A,B}$ of the two centrals spin of a chain of $L=12$ for the model in \eqref{eq:Ham1/2}. Panel a), $\rho(1,2)$ (see text) is discontinuous along the transition and $\rho(1,1)$ (panel b)) has a spike. In the second row, $\rho(1,2)$ and $\widetilde{\partial{B_x}}\rho(1,2)$ (derivative normalized by its minimum), (panel c) and d) respectively) obey a universal function when plotted against $\kappa_1$.}
\label{fig:elements}
\end{figure}
In this work, we have analyzed the behavior of bipartite entanglement measures in diverse 1QPTs. We have focused in 1QPTs which are influenced by a nearby 2QPT. In the first model we have considered, the spin-1/2 Ising model with longitudinal field, the concurrence, $\mathcal{C}$, shows always a continuous behavior along the 1QPT,  with a spike signalling a discontinuous first derivative.  Following \cite{Lidar04}, a singularity in the first derivative of $\mathcal{C}$ should signal a 2QPT given that this singularity comes strictly from the elements of $\rho^{ij}$, and not from the computation of $\mathcal{C}$ itself. In fact, in Ref.~\cite{JVidal04}, the authors compute $\mathcal{C}$ along a 2QPT and obtain  a similar behavior to our results in Fig.~\ref{fig:concu}. 
To determine the origin of this unusual behavior, in Fig.~\ref{fig:elements}, we look into the elements of $\rho^{A,B}$ of the two central spins. In the first column, we plot two of these elements as a function of $B_x$. The range of $B_x$ values used, is such that $|\kappa_1|\gg1$ avoiding, thus, the finite size effects described in Sec.~\ref{Section:Spin1/2}.
The first plotted element, $\rho(1,2)=\bra{\uparrow\uparrow}\rho^{A,B}\ket{\uparrow\downarrow}$, is discontinuous along the transition while the second, $\rho(1,1)=\bra{\uparrow\uparrow}\rho^{A,B}\ket{\uparrow\uparrow}$, has a spike signalling a singularity in the first derivative. In the second column we show that both singularities become smooth if we look inside the range $\kappa_1\sim 1$. We have added data for different $L$ and $B_z$ to show that these singularities fulfill the scaling for 1QPT described in \cite{Campostrini14}. When computing $\mathcal{C}$, the discontinuities present in elements such as $\rho(1,2)$ cancel out, but the spikes present in elements such as $\rho(1,1)$ do not.
As a result, $\mathcal{C}$ has a singularity in the first derivative as would happen in a 2QPT and it is precisely $\partial_{Bx}\mathcal{C}$ the quantity which fulfills the scaling ansatz and not $\mathcal{C}$ itself. Therefore, in this case, a non-analyticity in the derivative of the concurrence/negativity (given that the concurrence/negativity are continuous analytic functions), is not a sufficient condition to signal a 2QPT as it was conjectured, \cite{Lidar04}. The FSS shown in panel c) of Fig.~\ref{fig:concu} enables us to link the spike in $\mathcal{C}$ to a 1QPT despite that, at first sight, this behavior is the one expected for 2QPTs.\\
We have then applied our analysis to the spin-1 XXZ chain with on-site anisotropy. This system has several 1QPTs and we have found 4 patterns for $\mathcal{N}$ along these transitions. These patterns are: a discontinuity, a smooth slope, a dip (with a singularity in the first derivative) and a constant $\mathcal{N}$. The discontinuous behavior, which is the one expected for 1QPTs, is observed in the ferromagnetic/large $D$ transition where no finite size effects are present, even for small systems. For the large $D$/N\'eel transition, when crossing close enough to the tricritical point, a smooth slope in $\mathcal{N}$, instead of a discontinuity, appears due to finite size effects. Hence, one has to be careful when making  a one-to-one connection between 1QPTs and discontinuous bipartite entanglement for finite systems, especially when a 2QPT is close. The FSS for 1QPTs can link this smooth slope to a discontinuity in the thermodynamic limit. For the N\'eel/Anti-N\'eel, a dip in $\mathcal{N}$ is observed signalling a singularity in the first derivative. This behavior has the same origin as in the spin-1/2 case and, therefore, it is the derivative of $\mathcal{N}$ the one which fulfils the FSS for 1QPTs. Finally, we mention the ferromagnetic $\uparrow$/ferromagnetic $\downarrow$ transition where bipartite entanglement is constant and zero along the transition. In this case the discontinuities of the elements of $\rho^{A,B}$ trivially cancel out giving a constant and zero $\mathcal{N}$ and, thus, bipartite entanglement does not signal the QPT.\\
In conclusion, finite size effects become relevant when studying a 1QPT which is close to a 2QPT. Our main result is to have shown that $\mathcal{C}$ and $\mathcal{N}$ scale in a 1QPT as described in Ref.\cite{Campostrini14}.
We have observed that, close to a 1QPT, the discontinuities present in some elements of $\rho^{A,B}$ cancel out when computing the measures of bipartite entanglement but that the singularities present in the first derivative of some other elements do not. In this controversial example, where $\mathcal{C}$ has a spike at the critical point showing the apparent behavior of a 2QPT, the finite size scaling can unequivocally determine the order of the transition.\\
\textit{Acknowledgements}.  We acknowledge financial support from the Spanish MINECO projects FIS2013-40627-P, FIS2016-86681-P and the Generalitat de Catalunya CIRIT (2014-SGR-966). CC wishes to thank the EPSRC for support.\\
\bibliographystyle{apsrev4-1}
\bibliography{biblio_1OPT.bib}
\end{document}